\documentclass{elsart}

\usepackage{graphicx}
\usepackage{amssymb}

\begin{document}

\begin{frontmatter}

% Title, authors and addresses

% use the thanksref command within \title, \author or \address for footnotes;
% use the corauthref command within \author for corresponding author footnotes;
% use the ead command for the email address,
% and the form \ead[url] for the home page:
% \title{Title\thanksref{label1}}
% \thanks[label1]{}
% \author{Name\corauthref{cor1}\thanksref{label2}}
% \ead{email address}
% \ead[url]{home page}
% \thanks[label2]{}
% \corauth[cor1]{}
% \address{Address\thanksref{label3}}
% \thanks[label3]{}

\title{How fair is an equitable distribution?}

% use optional labels to link authors explicitly to addresses:
% \author[label1,label2]{}
% \address[label1]{}
% \address[label2]{}

\author[Kiel]{Elena Ram\'irez Barrios }
\author[Bremen]{Juan G. D\'iaz Ochoa}
\author[Mainz]{Johannes J. Schneider}
\address[Kiel] {Department of Economics,
Johannes Gutemberg University, D-55099 Mainz, Germany}
\address[Bremen]{Institute of Theoretical Physics,
University of Bremen, Otto Hahn Allee, D-28359 Bremen, Germany}
\address[Mainz] {Institute of Physics,
Johannes Gutenberg University, Staudinger Weg 7, D-55099 Mainz, Germany}

\begin{abstract}
Envy is a rather complex and irrational emotion. In general, it is very difficult to obtain a measure of this feeling, but in an economical context envy becomes an observable which can be measured. When various individuals compare their possessions, envy arises due to the inequality of their different allocations of commodities and different preferences. In this paper, we show that an equitable distribution of goods does not guarantee a state of fairness between agents and in general that envy cannot be controlled by tuning the distribution of goods.
\end{abstract}

\end{frontmatter}

% main text
\section{Introduction}
Envy is commonly defined as a feeling with a negative character that affects the social relationships making it rather complicated to define and, therefore, to measure. In economics, this envy relation can be established as a phenomenon that emerges after interpersonal comparisons between agents. From this point of view, envy can be considered as an economical observable that can be measured. We have developed a model where we analyze the emergence of envy in a network of agents when there is an allocation of goods. Each agent has a list with individual preferences in mind that restricts his/her choices. The satisfaction level at the end of the allocation process is measured according to the restrictions imposed by their preferences. This model could be applied in behavioral economics \cite{Kahneman},\cite{Arthur} and potentially in physics, in particular measurement problems, which are related to conditional probabilities \cite{Dolev},\cite{Kleidon}.

Previous investigations about similar systems are based on the development of a single model that describes the dynamics between agents as the dynamics of a network where the agents are located on its nodes \cite{Barabasi}\cite{Kirman}. An equivalent concept has been used by Donangelo et al.\ to model a network of trading agents as an interaction rule. Here the exchange between agents consists of goods as well as information that can be quantified \cite{Donangelo}. On the other hand, interpersonal relations have been studied in markets with a finite number of agents and a finite number of goods in economics without production, see e.g.\ Schmeidler et al.\ \cite{Schmeidler}. This paper exhibits a fundamental definition of fairness, depending on some trading properties of the agents and prices, without considering the quality of goods or a fairness index. 

Our motivation is to investigate interpersonal relations between the agents and to understand how the market motivates the evolution of these relations. We do not model envy as a network of trading agents interacting via envy relations, but we analyze the behavior of the agents as a function of parameters of allocation and the form in which agents compare their goods. Because emotions cannot be defined in a unique way, we develop a model that uses envy as a factor that modifies the conformation of a network based on information exchange. This information exchange is determined by the amount of envy. We suppose that an unidirectional exchange of information (each agent only observes his/her neighbors) is made through perfect channels, i.e., there is no noise or some other disruption in the transmission of information between agents. The existence of these channels ensures perfect comparisons, supposing that each agent has access to the content of information, i.e., the kind and number of goods assigned to other agents. This situation can be imagined as a group of agents with webcams, i.e., instruments to observe what other agents have, making them able to compare their actual situation with the situation of the other agents whom they can watch.
 
A maximized allocation, a situation in which each agent is reaching the best state with his/her endowment (see e.g.\ \cite{Feldman}), is called 'pareto' efficient. Otherwise, the individuals search for a larger allocation with a larger welfare, motivated by the possibility to get a larger utility level reflected in their interpersonal comparisons. Given the diffuse definition of the notion of envy, we present some fundamental concepts and theory from an economical point of view in Sec.\ \ref{sec2}. In Sec.\ \ref{sec3}, we explain the fundamental schemes and ideas in the formulation of our model. In Sec.\ \ref{sec4}, we propose our main results. Section \ref{sec5} is devoted to the main conclusions obtained in this investigation.
\section{Fundamental economical concepts\label{sec2}}
In economics, the first concept of equity, as no-envy, is due to Foley \cite{Foley}, introducing the concept of envy-free allocations: an allocation is equitable (or envy-free) if no agent prefers the bundle of another agent to his own. In this case we can say, there is a situation with fairness defined as no-envy.

The simplest problem of fairness is that of dividing a homogeneous commodity among a set of agents with equal claims on it. In this case, equal division (or equal income situation) is clearly the appropriate solution. If we want to have efficiency of this type of allocation and preserve its property of symmetry, the concept of fairness must be redefined. We define this equilibrium state as the moment when each person chooses the most preferred bundle in his/her budget set, and the choices exhaust the available supply.

An allocation is fair if it is envy-free and efficient \cite{Varian_I}. An allocation in a walrasian equilibrium with equal income is fair in this sense, but the converse no longer holds at all. A walrasian equilibrium is defined as a state where the aggregate excess demand (sum of all individual demands minus sum of all individual supplies) is equal to zero. So, if a bundle $B_i^\prime$ is preferred by agent $i$ to the bundle $B_i$ he/she currently possesses then the excess demand is different from zero. Hence, when the agents have equal initial endowments and equal possibilities in the market, they can easily reach their maximum utility getting a fair trade \cite{Schmeidler}.

A distribution of goods is said to be envy-free when no one prefers anyone else's bundle of resources to his/her own. The suggestion here is that envy is not the psychological motivation for the concern with equality, but rather that, where a distribution in fact produces envy, this is a reason to doubt the fairness of the distribution. But envy in this context is a technical term for any situation in which someone prefers another's bundles of goods, and does not refer to the emotional syndrome with which this envy is concerned.  
\section{Model\label{sec3}}
For the construction of our model, we need three basic elements: the first one is a set of various goods of $K$ different kinds located in a depot. For each kind $k$ of goods, a specific number $G(k)$ of goods exists. Our second element is a set of agents. There are $N$ agents in our model. Each agent $i$ has a preference list $P_i$ of his/her preferred goods. This preference list $P_i$ can be coded as a permutation of the numbers $1,\dots,K$ with $P_i(1)$ being the good being most important for agent $i$, $P_i(2)$ being the good second important, and so on. The third element is a set of individual 'baskets' $B_i$ in which the agents can deposit their goods after picking them up from the set of goods in the depot. $B_i(k)$ thus denotes the number of goods of kind $k$ the agent $i$ has got in his/her basket. After all the goods have been picked up by the agents at the depot, the following equation holds
\begin{equation}
\sum_{k=1}^K G(k) = \sum_{i=1}^N \sum_{k=1}^K B_i(k) + R
\end{equation}
with $G(k)$ being the number of goods of kind $k$ in the depot before the first agent has entered the depot to take some goods into his/her basket. $R$ are the remaining goods in the depot that any body wants.
%======================================================================
\begin{figure}[th]
\begin{center}
\includegraphics[clip,width=1\textwidth]{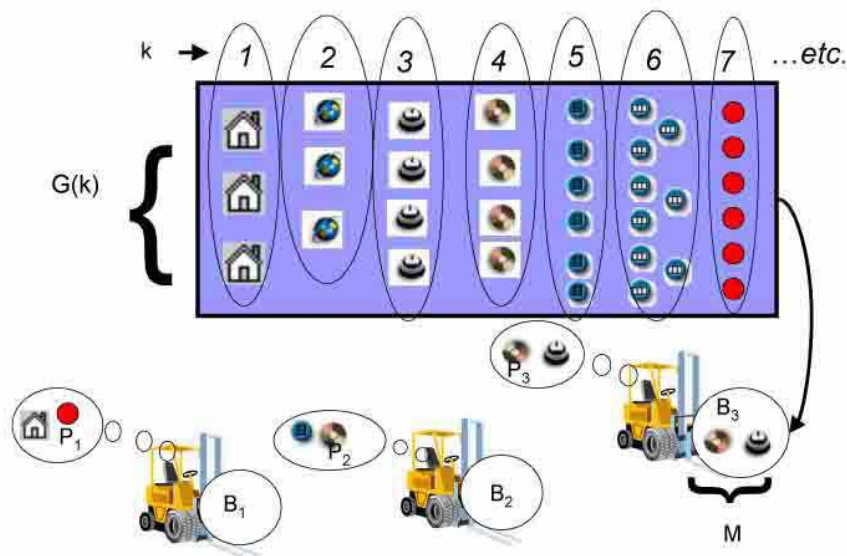}
\end{center}
\caption{System of agents in a depot, in which there is a predetermined amount of goods. Each agent has already a preference list $P_i$ in mind and is able to take goods according to his/her preference list and put them into his/her basket $B_i$. There are $K$ different kinds of goods, $G(k)$ is the number of goods available for each kind. Each agent is allowed to take a maximum number of goods $M$.}
\label{GOODS_DISTR}
\end{figure}
%======================================================================

According to their individual preferences, the agents search for goods of their highest ranked preference in the available set of goods. Every agent is allowed to take an overall number $M$ of goods. If the searched good of the highest preference is no more available, the agent starts to collect the good second on his/her preference list, and so on. Each agent takes the goods from the set of goods into his/her basket. This situation is shown in Fig.\ \ref{GOODS_DISTR}: several agents enter the depot in a random order. The first agent according to this randomly created queue takes $M$ stored goods (for example motorbikes, computers, etc.) into his/her basket. Then he/she leaves the depot and the second agent in the queue enters the depot. He/she also takes $M$ goods from the depot, selecting them according to the preference list he/she has in mind, and leaves the depot. This is repeated until the $N$ th agent takes the remaining $M$ goods. The overall number of goods is just sufficient for the agents in our model, i.e.,
\begin{equation}
N \times M + R= \sum_{k=1}^K G(k) .
\end{equation}
There is no production or destruction of goods. The number of agents also remains constant. So, when the agent takes a good into his/her basket, there is a depletion of one good in the depot.
%======================================================================
\begin{figure}[th]
\begin{center}
\includegraphics[clip, width=0.9\textwidth]{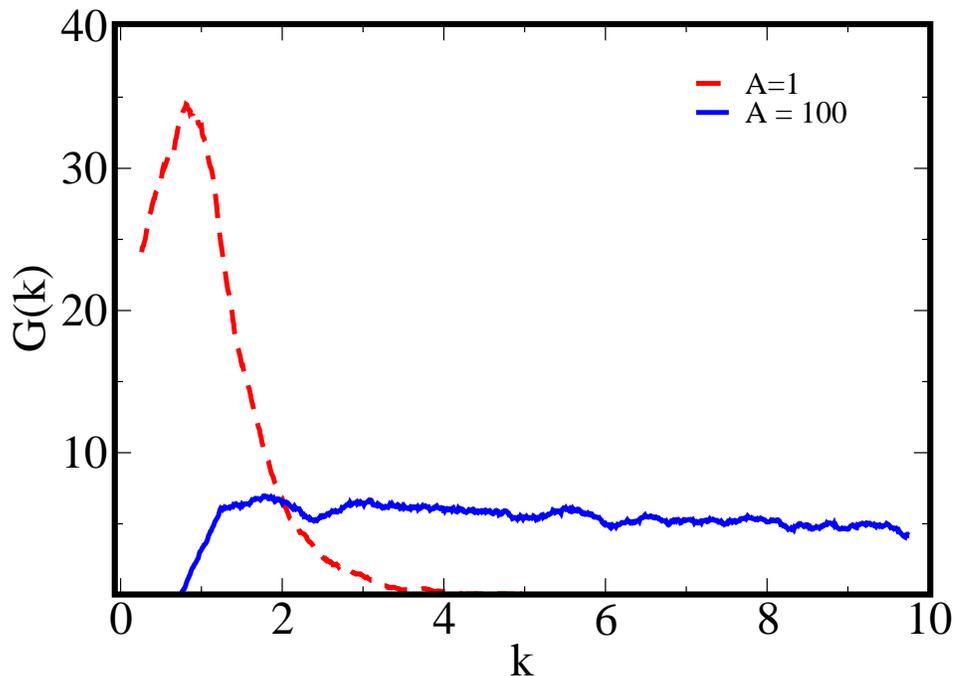}
\end{center}
\caption{Plot of the distribution of goods for two different distribution amplitudes. $k$ denotes the kind of goods and $G(k)$ the number of goods of kind $k$. $A$ is the parameter associated to the distribution amplitude of goods. }
\label{DISTR}
\end{figure}
%======================================================================

In a first approach, we keep the number of goods $G(k) \equiv G$ equal for each kind of goods in the depot in order to analyze i) fairness in equitable distributions and ii) the dependence between our parameters. Later on, we will introduce differences in the number of available goods. Following the rule of Knuth, we define the distribution of the number of goods $G(k)$ according to the Gamma distribution  
%%JJS Formel scheint nach gnuplot falsch zu sein
\begin{equation}
G(k)[A] = \frac{(x-k_{0})^{A-1}e^{-(x-k_{0})}}{\Gamma(A)},
\end{equation}
where  $k_{0}$ is the index of the good around which the distribution is peaked and $\Gamma(A)$ is the gamma function \cite{Mathworld}. In this equation the parameter $A$ is the amplitude of the Gamma distribution, i.e. the variation of this parameter produces a more or less uniform initial distribution of goods in the depot. The Gamma distribution has been used in a variety of settings, including the income distribution and production functions \cite{Greene} and for this reason it is a first choice for introducing it in our model. An example of the applied distribution is shown in Fig.\ \ref{DISTR}. (Furthermore, the Gamma distribution approaches the Gaussian distribution for large numbers, but does not exhibit some unwanted properties of the Gaussian distribution, like some finite probability for negative numbers. This is a very advantageous property of the Gamma distribution in comparison to the Gaussian distribution.)
%======================================================================
\begin{figure}[th]
\begin{center}
\includegraphics[clip,width=1\textwidth]{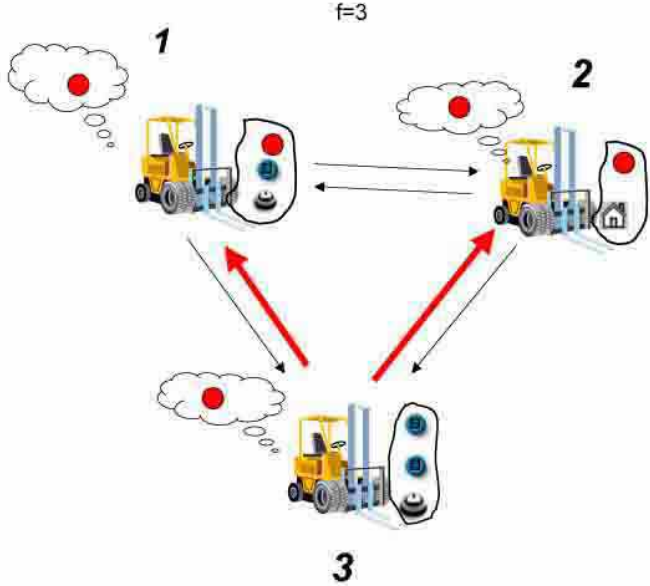}
\end{center}
\caption{Schema representing, how envy emerges among the agents $i=1, 2, 3$ with similar preferences (inside the clouds). The acquired bundle $B_i$ lies in front of each agent. A light arrow from some agent $i$ to an agent $j$ represents that agent $i$ looks into the basket of agent $j$. The bold arrow represents the bearing of the envy relation, because agents 1 and 2 acquired their desired goods but agent 3 did not. $f = 3$, where $f = 3$ represents the number of comparisons in the system.}
\label{ENVY_REL}
\end{figure}
%======================================================================

After the distribution of goods, the agents are allowed to take a look into the baskets of some other agents and to compare their own goods with the goods of these other agents according to their own preference lists. Some agents are satisfied because they filled their baskets better according to their own preference lists, while other individuals notice that the baskets they look into are filled with goods which are on top of their own preference lists but which they failed to get. In this case, envy emerges. This situation is shown in Fig.\ \ref{ENVY_REL}. 

Let us now define more clearly when envy occurs: two agents $i$ and $j$ are randomly chosen. They are allowed to look into the baskets of each other. Of course, they first check for the goods on top of their preference lists. Now if agent $i$ sees that $B_j(P_i(1))>B_i(P_i(1))$, i.e., that agent $j$ has more of the good on top of his/her preference list than himself/herself, then agent $i$ feels a strong surge of envy. Contrarily, if $B_j(P_i(1))<B_i(P_i(1))$, i.e., if agent $j$ has less of the good on top of the preference list of agent $i$, then agent $i$ is satisfied. In the case that $B_j(P_i(1))=B_i(P_i(1))$, agent $i$ checks for the second good on the preference list. Here again agent $i$ might get satisfied if $B_j(P_i(2))<B_i(P_i(2))$, envious if $B_j(P_i(2))>B_j(P_i(2))$, or willing to inspect the basket according to good $P_i(3)$ if $B_j(P_i(2))=B_i(P_i(2))$. This approach is repeated till 
%the last good if the goods are 
the baskets are equally filled with goods of higher priority. In the special case that both baskets are identical, no envy occurs. We can write this envy relation of the envy agent $i$ might feel towards agent $j$ formally as
\begin{equation} \label{Formel}
E_i(j) = \sum_{k=1}^K \Theta(b_j(P_i(k))-b_i(P_i(k))) \times
\prod_{l=1}^{k-1} \delta(b_i(P_i(l)),b_j(P_i(l)))
\end{equation}
with the Heaviside function
\begin{equation}
\Theta(x)=\cases{1 & if $x>0$ \cr 0 & otherwise\cr}
\end{equation}
and the Kronecker symbol
\begin{equation}
\delta(x,y)=\cases{1 & if $x=y$ \cr 0 & otherwise\cr}.
\end{equation}
Thus, one has
\begin{equation}
E_i(j)=\cases{1 & if agent $i$ feels envy towards agent $j$ \cr
0 & otherwise\cr} .
\end{equation}
Of course, when determining $E_i(j)$ in our simulations, we do not use Eq.\ (\ref{Formel}) but the approach described above 
%as this approach 
needs much less calculation time. %Please 
Note that the addend for $k=1$ in Eq.\ (\ref{Formel}) is welldefined as the empty product $\prod_{l=1}^0 \dots$ has a value of 1.

Thus, after the agents have taken the goods from the depot, we create pairs $(i,j)$ of agents who are allowed to look in each other's basket. The determination of the goods at the depot was performed in a completely rational manner by the agents. But now they are afraid that some other agent might have gotten some better goods out of the depot that himself/herself according to the own preference list. Here the behavior of the agents becomes irrational. Thus, we decide here not to give the agents some elaborate strategies at which other agents to look for baskets which are filled in a better way, i.e., with the more desired goods, but we implant the simple thought in their brains to randomly select other agents and look into their baskets. There is no particular difference between the agents except the contents of their baskets and their preference lists. Due to this equality of the agents and their irrationality, we think that it is appropriate to choose the pairs of agents completely at random.

At the same time when agent $i$ looks into the basket of agent $j$, agent $j$ looks into the basket of agent $i$. Due to the usually different preference lists $P_i$ and $P_j$ of agents $i$ and $j$, four different cases can occur: either none of them feels envy, as both of them are satisfied due to their different preference lists, or exactly one of them feels a strong surge of envy, whereas the other one feels satisfied, or both are envious towards each other.
%=========================================================================
\begin{figure}[bp]
\begin{center}
\includegraphics[clip,angle = -90,width=1\textwidth]{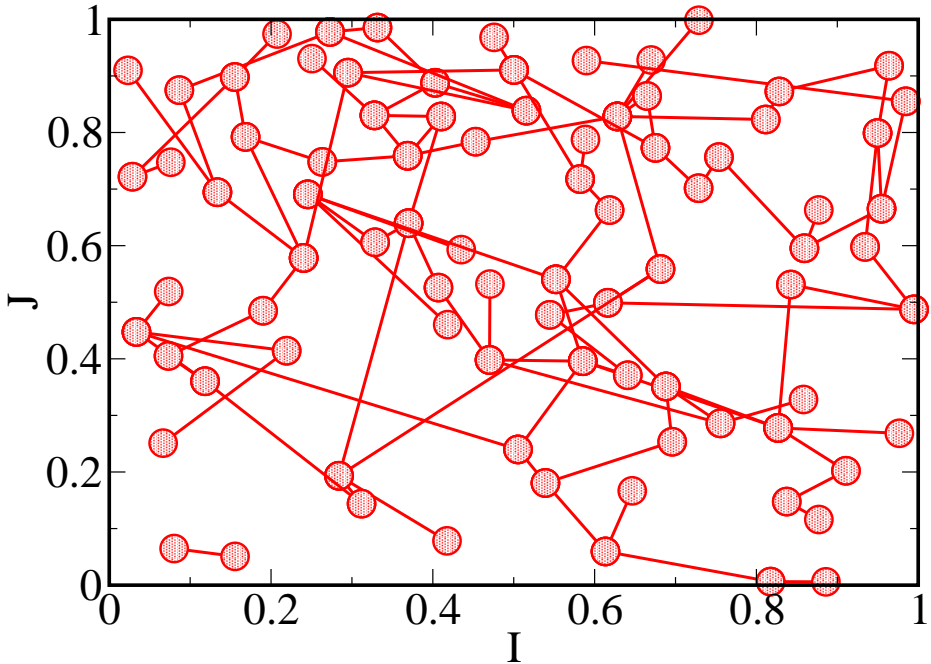}
\end{center}
\caption{Snapshot of a random network of interpersonal comparisons in a system of envious agents. The nodes in the network represent the agents. The edges between the nodes symbolize the relations that pairs of agents are allowed to look into the baskets of each other. Please note that the spatial arrangement of the agents does not play any role here.} 
\label{fig1}
\end{figure}
%=========================================================================

Each agent is allowed to make ``visual contact'' with $f$ other agent on average and to look into their baskets. The random selection of pairs of agents leads to a random network \cite{Bela}, as shown in Fig.\ \ref{fig1}. Please note that all edges in this network are undirected, i.e., if there is an edge from agent $i$ to agent $j$, also the edge from $j$ to $i$ exists. This random network can simply be described via a symmetric edge matrix $\eta$ with
\begin{equation}
\eta(i,j)=\cases{1 & if an edge between $i$ and $j$ exists \cr
0 & otherwise\cr}.
\end{equation}
After the edges have been chosen, we can define a Hamiltonian $E$ for this network of agents, summing up the amount of envy occurring in this network:
\begin{equation}
E=\sum_{i=1}^N \sum_{j=1}^N \eta(i,j) \times E_i(j)
\end{equation}
Note that the value of $E$ is usually different from the number $N_E$ of envious agents, which is given by
\begin{equation}
N_E=\sum_{i=1}^N \Theta\left(\sum_{j=1}^N \eta(i,j) \times E_i(j)\right) ,
\end{equation}
as a dog in the manger can feel envy towards more than only one other agent.

An envy network is created due to the assignment of goods to the baskets of the agents, who make comparisons based on their own preferences, and emerges only when the system is out of a Walrasian equilibrium, i.e. when there is a bad allocation of goods. Our model is not a typical optimization problem, in which the global optimum of a proposed pay-off function (free energy for instance) has to be found. When agents try to find a solution which is optimal for themselves, this solution, which is called Nash equilibrium, might not be a global optimum of the whole problem \cite{Johannes}. Our pay-off function here, which determines the amount of envy, is a measure of how far the system is from the equilibrium state. So we analyze the connectivity dependence on the assignation of goods and the number of agents that express envy. 

The fairness state, and not the topology of the network, is the main problem in this investigation \cite{Pastor}. We have particular interest in the measurement of the number $N_E$ of agents with envy. For this reason, envy is rescaled in our model by the total number of edges in the network divided through the total number of nodes (agents). There is a fairness state when there are no connections between the nodes of the random network, as then $E =0$. Otherwise, we recognize the emergence of an envy state. 

This network represents the bidirectional exchange of information between agents, i.e., each agent looks into the basket of the other agent searching for goods of his/her own preference. Because they can look at but not remove goods from the basket of the other agent or exchange goods, the envy relation due to interpersonal comparisons cannot be resolved.

Is it possible to improve the fairness by increasing the amplitude of the number of goods? At a first glance, this is a plausible way to get a very small envy network between the agents. We want to probe this hypothesis making a variation of the amplitude of the distribution of goods $A$.

According to the conventional definitions of fairness, an allocation is fair when it is also symmetric \cite{Kolm}. Theoretically, when an asymmetry is presented two different states are generated, namely inferiority, because an unsatisfied agent wants the object or good that the other agent has (object of envy) and, on the other hand, superiority, because an agent possesses a good that the other one wants \cite{Rawls}.

%%JJS Diesen Absatz verstehe ich leider nicht.
%Then we require the number $f$ of times each agent looks at one of his/her neighbors, and the number $K$ of different kinds of goods, as the main two variables of our model (the last one determines the size of the system, as both the vector length of the preference lists and the maximum amount of comparisons an agent performs while looking into one basket are given by $K$). The number $G(k)$ of goods of each kind is a parameter of the initial allocation of goods, previous to the distribution of the goods among the agents, that determines how equitable this distribution is. The effect of the homogeneity in the distribution of goods is analyzed with the introduction of the additional parameter $A$ for the amplitude of a nonuniform initial distribution of goods. Finally the parameter $M$ corresponds to the allocation of goods in the baskets of the individual agents. 

%The two main observables of our system are $E$, the number of times in which envy is expressed during the various comparisons, and the number $N_E$ of envious agents. The advantage of our approach is the possibility to measure these observables and thus the potential comparison of our simulation results with findings obtained in the real world using interviews with real human agents but also studies of apes and other animals which can express feelings and creatures in aritifical worlds.
%======================================================================
\begin{figure}[th]
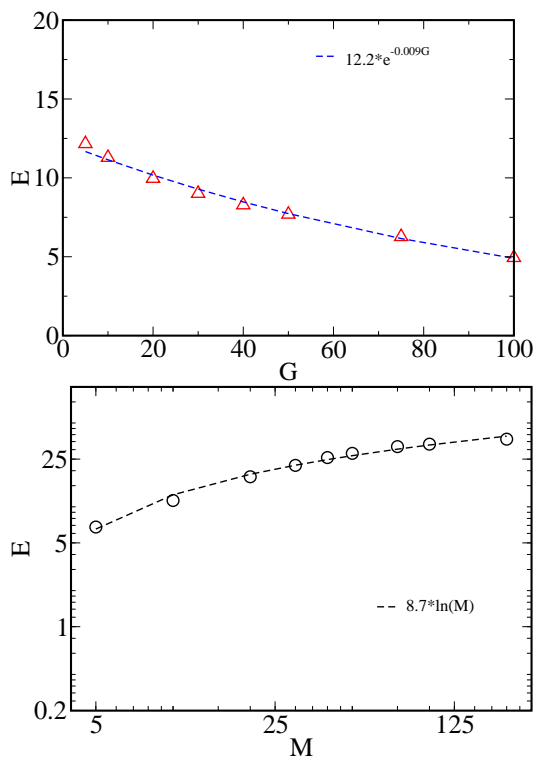

\begin{center}
\includegraphics[clip,width=0.5\textwidth]{Neid_vs_NG.eps}
\includegraphics[clip,width=0.5\textwidth]{Neid_vs_MaxNG.eps}
\end{center}
\caption{Envy $E$ as a function of $G$ (top) and $M$ (bottom). Each point was computed for distributions with maximal ten units of each kind of goods.}
\label{Neid_vs_NG}
\end{figure}
%======================================================================

\section{Results\label{sec4}}
We performed simulations with $N=100$ agents and $K=100$ different kinds of goods.
%%JJS Aber N*M war doch die Gesamtzahl, die fest vorgegeben ist.
The total number of goods in the system is determined in a random way governed by the Gamma distribution. Each agent can have a look into the basket of the neighboring agents. Given that the system is closed (there are no changes in the numbers of goods and the preferences remain the same) then the system must reach an equilibrium state in the distribution of envious agents for $f\rightarrow \infty$.

In a first set of simulations, we use $G(k)\equiv G$ for all kinds $k$ of goods. The envy $E$ as a function of $G$ is presented in Fig.\
\ref{Neid_vs_NG}.
%%JJS: Um das zu sehen, sollten Kurven fuer verschiedene f rein.
%%JJS: Aber indem G veraendert wird, muss doch auch M veraendert werden, oder?
The envy relation grows according to the increment of the number $f$ of contacts assigned to each agent. For large numbers of allowed comparisons, this result can be fitted with an exponential function, suggesting that the envy distribution corresponds to some kind of Boltzmann distribution in the network. This analogy makes it possible to relate $G$ to a kind of temperature of the system. 
%=====================================================================

\begin{figure}[ht]
\begin{center}
\includegraphics[clip, width=0.9\textwidth]{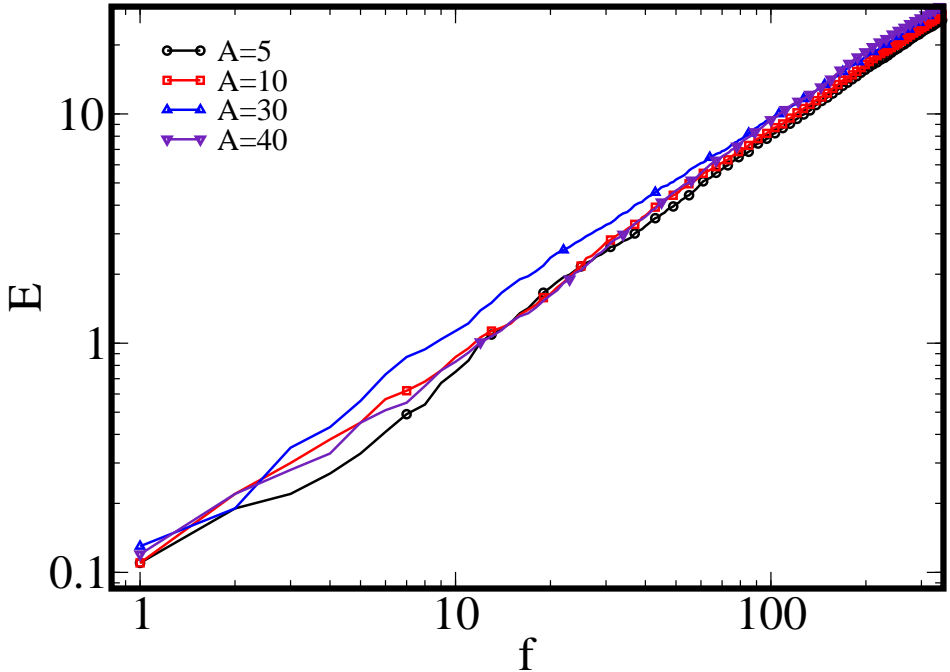}
\caption{General dependence on the envy between agents as a function of the number of interpersonal comparisons $f$. In this case we analyze the behavior of the relation between agents for different distribution amplitudes $A$. \label{ENVY_f}}
\end{center}
\end{figure}
%=========================================================================

A ''non walrasian equilibrium state'' is reached when the agents are allowed to explore all the baskets of the other agents. A fair state cannot be reached by increasing the supply of each kind of goods. This mechanism seems to be very clear in a society, where envy is the motor that ensures consumption. Innovation is related to the increment of the kind of goods supplied in the market. This increment shows that, contrary to  a stagnation, there is still a tendency to increase the consumption. In other words, innovation plays the role of a ''thermostat'' in a consumption society. 

The analysis of envy as a function of $M$ is also shown in Fig. \ref{Neid_vs_NG}. The increase of $M$ generates a logarithmic increase of envy. The shape of the fit does not depend on the number of comparisons allowed between agents. This is an empirical law that expresses how an increase in the allocated number of goods is not a guarantee for the improvement of fairness in a set of agents. Now the problem is to implement this situation in a nonuniform distribution of goods. 

Varian has demonstrated the existence of fair allocations \cite{Varian_I}. Also Kant called for better allocations that lower or suppress envy \cite{Kant}. As already mentioned above, we search for a fair allocation by controlling the distribution of goods through the amplitude parameter $A$. The expected result is a low envy relation by broadly distributed numbers of goods. These allocations must appear as fairer allocations.

In order to test the role of the distribution of goods for the initial allocation we fixed the number of kinds of goods in 100 and repeated our simulations for envy as a function of $f$ and $A$. Our computer simulations show a rather different scenario as the theoretical one: with a broad distribution, the envy in the network increases (see Fig. \ref{ENVY_f}). This implies, envy cannot be controlled by adjusting the distribution of goods. For the low number of allowed comparisons $f$, there is a convergence of the curves, computed as a function of the parameter $A$
%=====================================================================
\begin{figure}[ht]
\begin{center}
\includegraphics[clip, width=0.9\textwidth]{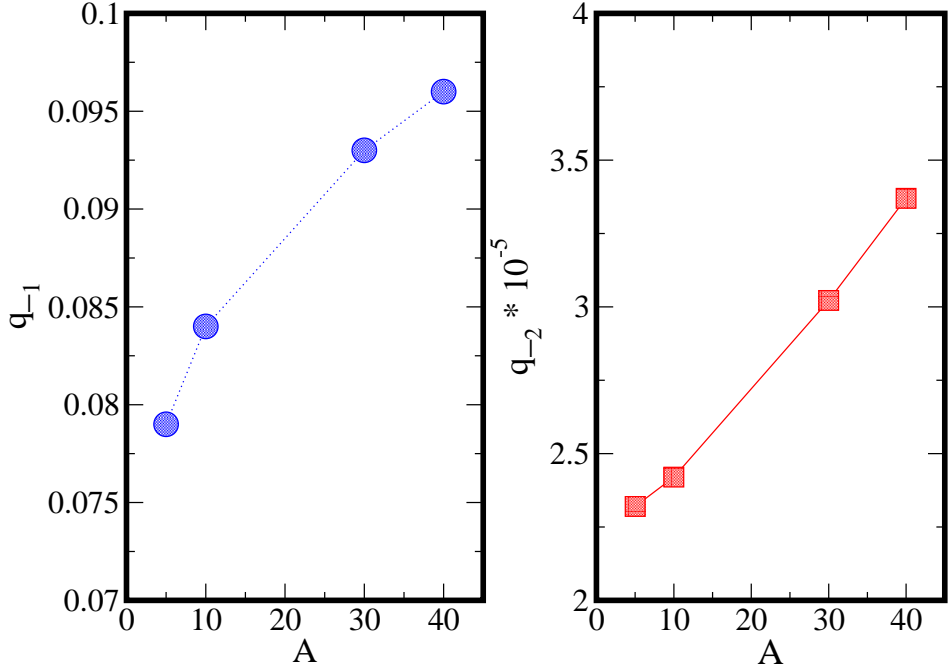}
\caption{Dependence of the parameters $q_{i}$ on the parameter $A$ for $i=1$ and $i=2$.\label{PARAM}}
\end{center}
\end{figure}
%=========================================================================

\begin{equation}
E = \sum_{i=1}^{n}q_{i}(fl)^{i}.
\end{equation}
Good agreement was found using polynomials of degree two. This means, this kind of system does not allow to become a preferred number of connections depending on the agent, a behavior commonly associated with random networks converging into free scale ones \cite{Barabasi}.

The parameters in the polynomial have dependence on the amplitude of distribution, i.e. $q_{i} = q_{i}(A)$. In particular the dependence of $q_{1}$ and $q_{2}$ on $A$ is shown in Fig.\ \ref{PARAM}. In both cases the Envy increase when the symmetry of the distribution is higher: The fairness is not improved by increasing the symmetry in the distribution of goods.  

\begin{figure}[th]
\begin{center}
\includegraphics[clip, width=0.9\textwidth]{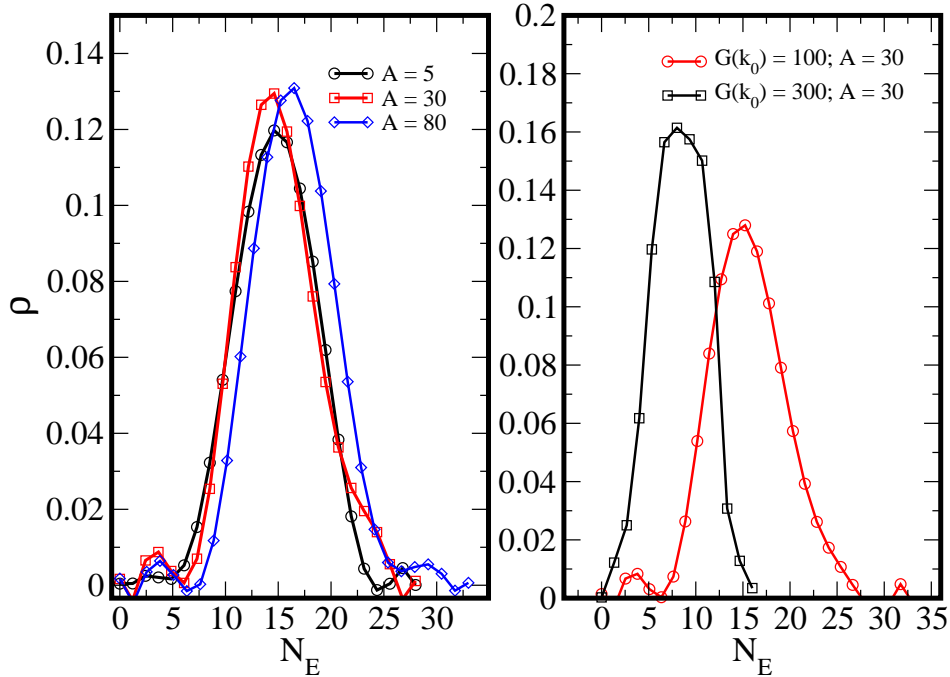}
\end{center}
\caption{Distribution of agents expressing envy. We analyze the behavior of the relation between agents for different distribution amplitudes $A$. $N_E$ is the number of agents expressing envy. $\rho$ is the density of this distribution.} 
\label{fig3}
\end{figure}

In order to test the role of the distribution of goods in the allocations, we repeated our simulations for envy as a function of $f$ and $A$. The results for this approach are shown in Fig.\ \ref{fig3}. In these simulations, the effect of the variation of $G(k)[A]$, on $N_E$ is analyzed. The number of envious agents is again small when the number of goods of each kind is large. So for $\langle G(k)[A]\rangle_k\approx 300$ the number of envious agents is roughly $7\%$ of the total number of agents, whereas for $\langle G(k)[A]\rangle_k\approx 100$ the fraction of envious agents is approximately $15\%$. If the number of goods is infinite and if there are only a few agents, then the number $N_E$ of envious agents is zero and thus $E$ is zero.
%=====================================================================
%=========================================================================

The variation of $A$ has an effect on the distribution function. This computation was made for $G(k_0)[A]=100$. For $A=5$ the simulations show a very asymmetric and narrow distribution of envious agents, while for $A=80$ the distribution of envious agents is slightly broader and shifted towards larger numbers $N_E$ of envious agents. Therefore, a uniform distribution increases the number of potentially envious agents. This result also means that a non uniform distribution boosts the number of envious agents with particular preferences. We can interpret these graphics under the light of a planified economy, for example, as a post-war system (in which the distribution of goods is rigorously controlled and in which the number of goods is comparable to the number of agents in the system) and a capitalistic system. In the first case, no matter how uniform the distribution of goods is, the probability to find an envious agent never vanishes. In the second case, the narrow distribution of agents shows the way how the initial distribution of goods can be adapted in order to satisfy the demand. 
\section{Concluding Remarks\label{sec5}}
The construction of models of agents with particular features, like societies, is often hard to implement\footnote{And set the problem presented in the Arrow's impossibility theorem}. Furthermore the comprehension of the emotional (subjective) component in a trading network is fundamental to develop strategies in order to understand economical and social systems. The model presented here is a static portrait in a state equivalent to an equilibrium state for a closed system where a set of agents with individual preferences is allowed to take objects from a depot with an initial distribution of goods. The results show, under which conditions the agents are not satisfied with the goods, they are able to acquire. Two equilibrium concepts arising, one from the economical theory and the other from statistical mechanics. These equilibria are required for the construction of the present model. From an economical point of view, this system is far from equilibrium (defined as a Walrasian one, where the sum of total demands must be equal to the sum of total supplies of the economy) when the agents develop interpersonal comparisons. Here, the number of interpersonal comparisons is equivalent to the magnitude of an envy field. 

From the physical point of view, the system reaches an equilibrium state (statistical equilibrium) when the agents are able to make enough interpersonal comparisons in order to explore the whole space of agents; after a large number of interpersonal omparisons it is possible to quantify a distribution of envious agents that does not change in time. Furthermore, the existence of a statistical equilibrium is tested by simulations, based on a Monte Carlo method.

The number of envious agents depends on the number of possible comparisons and the amount of available goods in the system. For very few visual contacts between the agents, the envy level is zero, and only after many interpersonal comparisons, i.e.\ after a long time, the system reaches a stable envy state (please note that we do not consider time directly, but the parameter $f$ of the number of comparisons can be interpreted as a measure of time). Also when the number of agents is small compared to the number of goods, the number of envious agents tends to be zero. These results imply a simple conclusion: agents with very good allocations and agents with no visual contact to other agents do not have any chance to develop envy. 

The control of the distribution of the number of goods is a way to control the amount of envy among the agents. However, the results obtained in the present calculations are quite contra-intuitive: a uniform distribution of goods does not decrease the number of envious agents.

This model appears like a caricature of a society of agents with feelings. We show what could happen when the information of the preferences of the agents is accessible and the interaction is fixed. In contrast, in a real scenario all agents can develop particular strategies in order to get the goods they want and to improve their welfare, formally represented by means of their utility functions. Furthermore, the opinion of an agent is not fixed and could change when it is exposed to information that changes his/her 'mind'. A good example is a magazine about mode and superstars, offering new goods that change the preferences of the people and the market supply. Therefore, an interesting perspective of this investigation could be the consideration of an evolutive panorama.

While agents remain in the system who would like to have the goods that the other agent at whom they look possesses, we show in the present results the impossibility to reach the desired and ideal Walrasian equilibrium state by controlling the distribution of goods in the depot, when the number of goods is limited. That means, an equilibrium in statistics implies a non Walrasian equilibrium. Models that far from equilibrium could be considered by introducing changes in the numbers of goods or agents in the system, or with dynamical preference lists for the agents, which change in time and thus force the agents to adjust the contents of their baskets by taking other available goods which are then on top of their preference lists, by which the Walrasian equilibrium is improved.

We acknowledge Dr. Takuya Yamano for his corrections and critiques of a preliminary version of the text, Dr.\ Rainer Diaz for his objections and interesting critiques about the economical aspects in the frame of the social sciences, Prof.\ Raul Donangelo for very useful remarks and the very interesting comments of three anonymous referees.

\bigskip

\bigskip

%\bibliography{paper}% Produces the bibliography via BibTeX.

% The Appendices part is started with the command \appendix;
% appendix sections are then done as normal sections
% \appendix

% \section{}
% \label{}

% Bibliographic references with the natbib package:
% Parenthetical: \citep{Bai92} produces (Bailyn 1992).
% Textual: \citet{Bai95} produces Bailyn et al. (1995).
% An affix and part of a reference:
%   \citep[e.g.][Ch. 2]{Bar76}
%   produces (e.g. Barnes et al. 1976, Ch. 2).

\end{document}